\documentclass[pdflatex,sn-mathphys-num]{sn-jnl}

\usepackage{graphicx}%
\usepackage{multirow}%
\usepackage{amsmath,amssymb,amsfonts}%
\usepackage{amsthm}%
\usepackage{mathrsfs}%
\usepackage[title]{appendix}%
\usepackage{xcolor}%
\usepackage{textcomp}%
\usepackage{manyfoot}%
\usepackage{booktabs}%
\usepackage{algorithm}%
\usepackage{algorithmicx}%
\usepackage{algpseudocode}%
\usepackage{listings}%

\theoremstyle{thmstyleone}%
\theoremstyle{thmstyletwo}%

\theoremstyle{thmstylethree}%

\begin{document}

\title[Article Title]{Full-field prediction for engineering-scale three-dimensional aircraft with multigrid-hierarchical learning}


\author[1,2,3,4]{\fnm{Yunfei} \sur{Liu}}
\equalcont{These authors contributed equally to this work.}

\author[1]{\fnm{Hao} \sur{Wang}}
\equalcont{These authors contributed equally to this work.}

\author[1]{\fnm{Yuhang} \sur{Qi}}

\author[1]{\fnm{Hao} \sur{Yue}}

\author[1]{\fnm{Dehong} \sur{Meng}}

\author[1]{\fnm{Wei} \sur{Li}}

\author[1]{\fnm{Rui} \sur{Wang}}

\author[4]{\fnm{Tiejun} \sur{Li}}

\author[2,3,4]{\fnm{Jie} \sur{Liu}}

\author*[1]{\fnm{Junwu} \sur{Hong}}\email{hjw128@sina.com}

\author*[2,3,4]{\fnm{Xinhai} \sur{Chen}}\email{chenxinhai16@nudt.edu.cn}

\affil[1]{\orgdiv{Computational Aerodynamics Institute}, \orgname{China Aerodynamics Research and Development Center}, \orgaddress{\city{Mianyang}, \postcode{621000},\country{China}}}

\affil[2]{\orgdiv{Laboratory of Digitizing Software for Frontier Equipment},\orgname{National University of Defense Technology},\orgaddress{\city{Changsha},\postcode{410073},\country{China}}}

\affil[3]{\orgdiv{National Key Laboratory of Parallel and Distributed Computing}, \orgname{National University of Defense Technology}, \orgaddress{\city{Changsha},\postcode{410073},\country{China}}}

\affil[4]{\orgdiv{College of Computer Science and Technology}, \orgname{National University of Defense Technology}, \orgaddress{\city{Changsha},\postcode{410073},\country{China}}}


\abstract{
High-fidelity computational fluid dynamics is essential for aerospace design, but engineering-scale simulations of practical three-dimensional aircraft remain computationally expensive. Learning-based flow-field initialization offers a complementary route to improve efficiency by reducing the numerical distance between the initial and converged solutions. In practice, existing deep learning approaches remain difficult to scale to engineering-scale three-dimensional flows because of the large parameter space and the intrinsic heterogeneity of multiscale flow regimes. Most studies therefore focus on two-dimensional problems, surface quantities or integral aerodynamic coefficients for three-dimensional aircraft, or full-field prediction only in simplified three-dimensional test cases with limited grid resolution. These settings either lack complete volumetric flow information, preventing physical consistency checks and coupling with downstream CFD solvers, or remain difficult to scale to engineering-scale aircraft flows with complex body-fitted boundaries, strong local gradients and million-scale grids. 
Here we propose MHLF, a multigrid-hierarchical learning framework for accelerating engineering-scale three-dimensional aircraft flow simulations while preserving high-fidelity numerical accuracy. MHLF combines a topologically consistent Geometric MultiGrid representation, which reduces the learning burden on engineering-scale grids, with a hierarchical strategy that captures regional flow heterogeneity during both prediction and subsequent CFD correction. Across three engineering-scale aircraft cases spanning Mach 0.15 to 6.0 and covering subsonic, transonic and supersonic regimes, MHLF accelerates convergence without sacrificing flow-field accuracy, achieving an efficiency improvement of 3 to 8 times over conventional initialization. Numerical results demonstrate that this method not only achieves the first large-scale, practical full-flow-field prediction for 3D aircraft within the CFD domain, but also establishes a solid foundation for future related research.
}



\maketitle

\section{Introduction}\label{sec:introduction}

Computational fluid dynamics (CFD) is central to aerodynamic design and optimization in aerospace engineering, providing reliable physics-based analysis at substantially lower cost than extensive physical testing. However, high-fidelity simulations of engineering-scale three-dimensional aircraft typically require millions of computational cells and repeated evaluations across design conditions, making CFD computationally expensive. Although CFD algorithms and solver technologies continue to advance, the cost of reaching convergence remains a major constraint on practical design cycles. This motivates complementary acceleration strategies that preserve the reliability of established CFD solvers while reducing the numerical distance from the initial state to the converged solution.

Deep learning has created new opportunities for flow-field prediction and CFD acceleration. Early studies mainly used regular-grid representations, treating flow variables as multichannel images. Convolutional neural networks enabled efficient prediction of two-dimensional steady laminar and RANS flow fields, demonstrating the feasibility of image-based learning for fluid problems \citep{guo2016convolutional,bhatnagar2019prediction,ribeiro2020deepcfd,thuerey2020deep}. Related ideas were extended to inverse design, where models learned mappings between target pressure distributions and airfoil geometries \citep{sekar2019inverse}. In three dimensions, voxel, octree and point-voxel representations reduced the cost of sparse volumetric learning \citep{riegler2017octnet,liu2019point}. These approaches advanced fast flow prediction, but regularized spatial representations can distort complex aircraft boundaries and near-wall flow structures in aircraft CFD \citep{vanska2025voxel}.

To improve geometric adaptability while retaining the efficiency of convolutional models, coordinate-mapping methods were introduced to transform irregular physical regions into regular reference regions \citep{gao2021phygeonet}. This strategy relaxed the geometric constraints of image-based flow prediction, but still relied on a structured reference representation. Subsequent work moved further towards graph structures and point-cloud representations. Graph neural networks showed strong representational capacity for flow problems with variable topology \citep{pfaff2020learning,belbute2020combining}. Point-cloud methods avoided explicit grid regularization and enabled learning directly from non-uniformly sampled geometric coordinates \citep{kashefi2021point,chen2024developing}. These developments improved the representation of complex geometries, but practical aircraft CFD imposes a more demanding setting.

Despite these advances, most learning-based aerodynamic prediction studies still reduce the difficulty of the problem through simplified settings. Existing work commonly focuses on two-dimensional fields, surface quantities or integral aerodynamic coefficients, or evaluates full-field prediction only in simplified three-dimensional cases with limited resolution \citep{bonnet2022airfrans,sabater2022fast,li2022deep,hines2023graph,shen2023deep,lei2024prediction,wang2025aerodynamic,catalani2024neural,rabeh2025benchmarking,nemati2024fast,xie2024fast,li2025fast,zuo2025flow3dnet}. These settings have demonstrated the potential of learning-based flow prediction, but they remain insufficient for practical aircraft CFD. They do not resolve two difficulties central to this setting, namely learning from large volumetric grids at engineering scale and representing flow regions with markedly different physical characteristics and convergence behavior.

The first difficulty is grid scale. Engineering-scale high-fidelity aircraft simulations typically rely on volumetric grids with millions of cells or more. Direct end-to-end learning on the original fine grid rapidly increases computational complexity, memory consumption and optimization difficulty. This issue is especially pronounced for graph and point-cloud networks as the number of nodes grows, affecting both training stability and computational efficiency \citep{lino2023current}. Global message passing on large graphs can also cause oversmoothing, weakening the representation of local high-frequency flow structures \citep{chen2020measuring}.
The second difficulty is cross-region flow heterogeneity. Practical three-dimensional external flows contain boundary-layer, inner-field and outer-field regions with different dynamics, grid scales, variable distributions and numerical stiffness. Boundary-layer regions contain steep gradients and high-frequency structures, whereas the outer field is smoother and varies more weakly \citep{zeng2025point}. A global representation compresses these distinct regimes into a single feature space, which can mask near-wall or shock-related structures and force multiscale features to compete within shared parameters. Because deep networks are biased towards low-frequency components, uniform modeling can reduce their ability to capture boundary layers, shocks and separated flows \citep{xu2019frequency,wang2021eigenvector}. Regional differences in variable distribution and convergence difficulty can also introduce gradient conflicts during joint training \citep{xiao2024mh,wang2021understanding}. Scalable full-field prediction for practical aircraft CFD therefore requires both a grid representation that reduces the learning burden and a flow-field organization that respects regional physical differences.

To address these challenges, we propose MHLF, a multigrid-hierarchical learning framework for accelerating engineering-scale three-dimensional flow simulations while preserving high-fidelity numerical accuracy. Rather than performing global end-to-end regression on the original fine grid, MHLF reformulates full-field prediction as a coupled process that combines low-cost learning on a topologically consistent coarse grid, region-aware modeling, high-resolution reconstruction and CFD correction. The framework constructs a coarse-grid representation while preserving the topology of the original CFD grid. This transfers the learning task to a lower-resolution but geometrically consistent representation space, reducing the computational complexity and optimization difficulty of large-scale flow-field modeling. The computational domain is then organized into boundary-layer, inner-field and outer-field regions according to flow-field characteristics. This partition allows the model to learn multiscale flow features under more balanced regional scales and variable distributions.
On this partitioned representation, MHLF uses Backbone--Branch Modulation and introduces a Region-Aware Residual Attention mechanism. These components adapt feature responses across different physical regions, mitigating feature masking and optimization conflicts under unified modeling. The predicted coarse-grid flow field is reconstructed at the original CFD resolution through multigrid prolongation \citep{brandt2011multigrid,brandt1977multi}, providing an initial flow fields for subsequent CFD correction. The same flow-field hierarchy is also used for numerical iteration scheduling, allowing different regions to employ distinct iterative strategies based on their flow characteristics and convergence behavior \citep{meng2026hierarchical}. In MHLF, multigrid representation provides the multiresolution learning and reconstruction pathway needed for large-scale flow-field prediction, whereas flow-field hierarchy provides the core mechanism for resolving cross-region physical differences. Together, they form an efficient full-field prediction framework for engineering-scale three-dimensional aircraft.

The main contributions of this work are as follows:
\begin{itemize}
    \item We propose a topologically consistent multigrid representation and reconstruction method for engineering-scale full-field prediction. This method converts high-resolution flow-field prediction into feature learning in a lower-cost coarse-grid space, and uses multigrid prolongation to provide initial flow fields at the original CFD resolution.
    \item We propose a flow-field hierarchy based on regional physical characteristics. By organizing the spatial flow field into boundary-layer, inner-field and outer-field regions, MHLF enables region-aware learning and solution strategies. During learning, hierarchical modeling is achieved through Backbone--Branch Modulation and Region-Aware Residual Attention. During CFD correction, region-specific scheduling improves numerical efficiency.
    \item To the best of our knowledge, this work represents the first full-field prediction for engineering-scale three-dimensional aircraft configurations across different Mach number regimes in the field of CFD. It has achieved remarkable results, marking a significant step in advancing deep learning-based flow prediction from laboratory research toward fully practical engineering applications, and opening a door for the comprehensive integration of deep learning methods into CFD. Building upon this work, flow-field prediction will no longer be confined to two-dimensional problems or simple three-dimensional geometries with basic flow structures; instead, it is poised for large-scale, extensive application to truly complex three-dimensional configurations and intricate flow regimes across the full velocity range.The numerical experiments in this paper demonstrate that the MHLF prediction framework not only showcases the immense potential of deep learning for flow-field prediction but also provides a solid foundation for future development.
\end{itemize}

\section{Results}\label{sec2}
We propose MHLF, a multigrid-hierarchical learning framework for accelerating engineering-scale three-dimensional flow simulations while preserving high-fidelity numerical accuracy. The central idea of this framework is to transfer learning across spatial resolutions through a topologically consistent multigrid representation, enabling feature learning in a low-resolution space and flow-field reconstruction at the original CFD resolution. At the same time, the flow domain is organized into boundary-layer, inner-field and outer-field regions according to flow-field characteristics. A Backbone--Branch Modulation architecture with a Region-Aware Residual Attention mechanism is then used for region-aware training. The predicted flow field provides a initial flow fields for subsequent CFD correction, substantially improving solution efficiency. The architecture and workflow of MHLF are shown in Fig.~\ref{fig:framework}. The multigrid coarsening and prolongation procedures are shown in Fig.~\ref{fig:multigrid}, and the prediction network is shown in Fig.~\ref{fig:partition-net}. Further details are provided in the Methods. Throughout the figures, the flow-field regions are denoted using a consistent color scheme, with orange representing the boundary-layer region, blue the inner-field region, and green the outer-field region.

To evaluate the effectiveness and computational efficiency of MHLF, we performed validation calculations on three canonical aerospace configurations using structured grids. The cases span $Ma=0.15\sim6.0$ and cover subsonic, transonic and supersonic regimes. The subsonic case is the half-span Trap Wing configuration from the first High Lift Prediction Workshop. The transonic case is the WBNP configuration of the DLR-F6 model. The supersonic case is the X-38 lifting-body configuration. These benchmarks are widely used in CFD verification and validation and involve different aircraft geometries, flow structures and grid scales, providing a practical test of both multigrid learning and flow-field hierarchy. Because of the large grid sizes, all simulations were performed in parallel. The comparisons show that the MHLF-based flow prediction and simulation strategy accelerates CFD convergence without sacrificing flow-field accuracy, achieving a threefold to eightfold efficiency improvement over conventional initialization and substantially outperforming the tested baseline methods. The following sections analyze the cases in order of increasing geometric complexity.

\begin{figure}[!htbp]
    \centering
    \includegraphics[width=\textwidth]{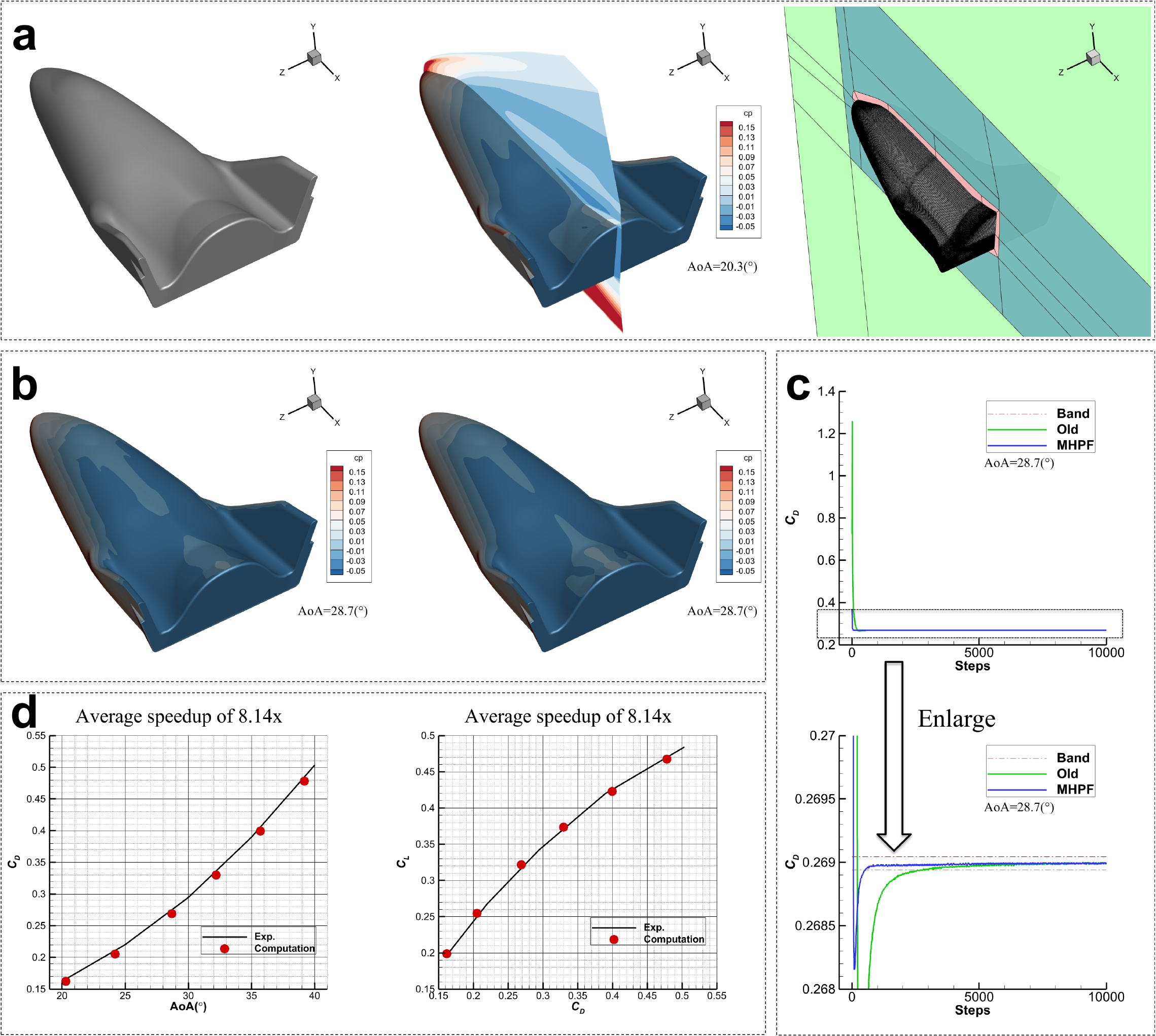}
    \caption{\textbf{X-38 computational geometry and comparison of computational results.}
a, Computational geometry, representative volumetric flow field and flow-field regions on grids at different layers. The computational grid contains 91 blocks and 7,412,224 cells, with $y^+\approx1$. b, Comparison between the predicted ( left ) and converge ( right ) surface-pressure contours at $\alpha=28.7^\circ$. The predicted flow clearly captures the main flow structures and closely resembles the converged solution. c, Comparison and local enlargement of drag convergence curves obtained using conventional CFD and MHLF at the representative angle of attack. The green line denotes conventional CFD, the blue line denotes MHLF and the red band denotes the convergence interval. The convergence curve obtained using MHLF has a much smaller oscillation amplitude and enters the convergence interval earlier, indicating a substantial increase in convergence speed. d, Comparison of drag coefficient and lift-to-drag ratio obtained using MHLF with wind-tunnel measurements. The close agreement supports the reliability of MHLF.}
    \label{fig:X-38}
\end{figure}

\subsection{X-38 case}

The X-38 case is a canonical supersonic CFD validation case released by NASA and derived from a real spacecraft program. It is a high-volume lifting-body configuration and serves as a standard model with typical supersonic flow features, including detached shocks and viscous interaction. The case is mainly used to assess the reliability and accuracy of CFD algorithms for supersonic aerodynamic prediction. Although the geometry is relatively simple, complex shock systems arise at high Mach numbers and large angles of attack, including the bow shock near the nose and leading-edge shocks near the wing. The interactions between these shocks and the body boundary-layer, together with large-scale flow separation on the upper wing surface and aft-body at high angle of attack, make the case challenging for CFD. The X-38 therefore represents a supersonic case with highly complex spatial flow structures and provides a stringent test for the proposed prediction model.

We first coarsened the original grid to obtain a training grid with $1/64$ of the cells in the original fine grid, and then performed numerical simulations on this grid. The training conditions were $Ma=6.0$ and $Re=3.10\times10^{6}$, with $\alpha=20^\circ \sim 40^\circ$ at intervals of $0.5^\circ$. The governing equations were the full Navier--Stokes equations, and the turbulence model was the two-equation SST model. This step incurred negligible computational cost, owing to the substantially reduced grid size relative to the original grid and the accelerated flow convergence caused by increased numerical viscosity after coarsening.

Based on these data, we trained MHLF on the coarse-grid flow fields. Training used the Adam optimizer with an initial learning rate of $1\times10^{-3}$ and momentum parameters $\beta_1=0.9$ and $\beta_2=0.999$. The learning rate was scheduled using StepLR and decayed by a factor of $0.8$ every $50$ epochs. All models were trained for $500$ epochs on an NVIDIA A100 GPU with $40\,\mathrm{GB}$ memory. After training, the model directly predicted the corresponding three-dimensional coarse-grid flow field for a given angle of attack. To more strictly evaluate predictive capability, all test conditions were excluded from the training angles. The selected comparison conditions were $\alpha=20.3^\circ$, $24.2^\circ$, $28.7^\circ$, $32.2^\circ$, $35.7^\circ$ and $39.2^\circ$. The computational geometry and prediction results are shown in Fig.~\ref{fig:X-38}.

\begin{table}[!htbp]
\centering
\caption{
Convergence performance and computational cost for the X-38 case.
}
\label{tab:X-38_steps_time}
\renewcommand{\arraystretch}{1.15}
\setlength{\tabcolsep}{5pt}
\begin{tabular}{lccccccccc}
\toprule
\multirow{2}{*}{Method} &
\multicolumn{6}{c}{Convergence steps $N_c$ at AoA $\alpha$ (deg)} &
\multirow{2}{*}{Total steps} &
\multirow{2}{*}{Total time (s)} \\
\cmidrule(lr){2-7}
& $20.3^\circ$ & $24.2^\circ$ & $28.7^\circ$ & $32.2^\circ$ & $35.7^\circ$ & $39.2^\circ$
& & \\
\midrule
Old       & 4512 & 4176 & 3180 & 3084 & 2508 & 2496 & 19956 & 31530.5 \\
PointCNN           & $\times$ & $\times$ & $\times$ & $\times$ & $\times$ & $\times$ & $\times$ & $\times$ \\
PointConv          & 4080 & 4032 & 3240 & 3096 & 2400 & 2328 & 19176 & 30298.1 \\
PointMLP           & $\times$ & $\times$ & $\times$ & $\times$ & $\times$ & $\times$ & $\times$ & $\times$ \\
PointNet++          & 5004 & $\times$ & $\times$ & $\times$ & $\times$ & $\times$ & $\times$ & $\times$ \\
PointNeXt          & $\times$ & $\times$ & $\times$ & $\times$ & $\times$ & $\times$ & $\times$ & $\times$ \\
PointTransformerV3 & 5868 & 5424 & 5016 & 4200 & 3660 & 3336 & 27504 & 43456.3 \\
\textbf{MHLF (ours)}               & $\mathbf{732}^{1}$ & $\mathbf{468}^{1}$ & $\mathbf{612}^{1}$ &
                     $\mathbf{1020}^{1}$ & $\mathbf{528}^{2}$ & $\mathbf{1056}^{3}$ &
                     $\mathbf{4416}$ & $\mathbf{3874.9}$ \\
\bottomrule
\end{tabular}

\vspace{2mm}
\begin{flushleft}
\footnotesize
The iteration strategy of the asynchronous iteration mode is slightly adjusted for different angles of attack. Superscripts 1, 2 and 3 denote the 10-3-1, 10-4-1 and 10-5-1 modes, respectively. In the 10-3-1 mode, the boundary-layer, inner-field and outer-field are iterated for 10, 3 and 1 steps, respectively, within each multigrid cycle. The other modes follow the same convention.
\end{flushleft}
\end{table}

Table~\ref{tab:X-38_steps_time} compares MHLF with conventional CFD and baseline models. For all computational conditions, three strategies were compared. Conventional CFD uses a uniform initial condition with synchronous iteration. The baseline strategies use initial conditions predicted by baseline models with synchronous iteration. MHLF uses its predicted initial condition with hierarchical asynchronous iteration. The six baseline models are PointCNN, PointConv, PointMLP, PointNet++, PointNeXt and PointTransformerV3, all of which are representative flow-field prediction models. In the table, AoA denotes the angle of attack, $N_c$ denotes the number of convergence steps determined by the convergence criterion described in Appendix 1, and Total steps and Total time denote the total number of computational steps and total computational time, respectively. The symbol $\times$ denotes calculation failure. The comparison shows that the baseline models generally have poor predictive capability for flow fields because they do not use hierarchical modeling and cannot effectively distinguish flow features across physical regions from macroscopic to microscopic scales. Most predicted initial conditions therefore do not accelerate convergence, and many even lead to calculation failure because the predicted flow variables deviate substantially from physical behavior. By contrast, the total number of computational steps required by MHLF over the six comparison conditions is only 22.1\% of that required by conventional CFD. Because asynchronous iteration also reduces the computational cost per step, MHLF achieves an average speed-up of 8.14 in this case, with the largest speed-up of 18.23 at $\alpha=24.2^\circ$. Given the flow complexity of the X-38 case, these results demonstrate the strong potential of MHLF.

\subsection{DLR-F6-WBNP case}

\begin{figure}[!htbp]
    \centering
    \includegraphics[width=\textwidth]{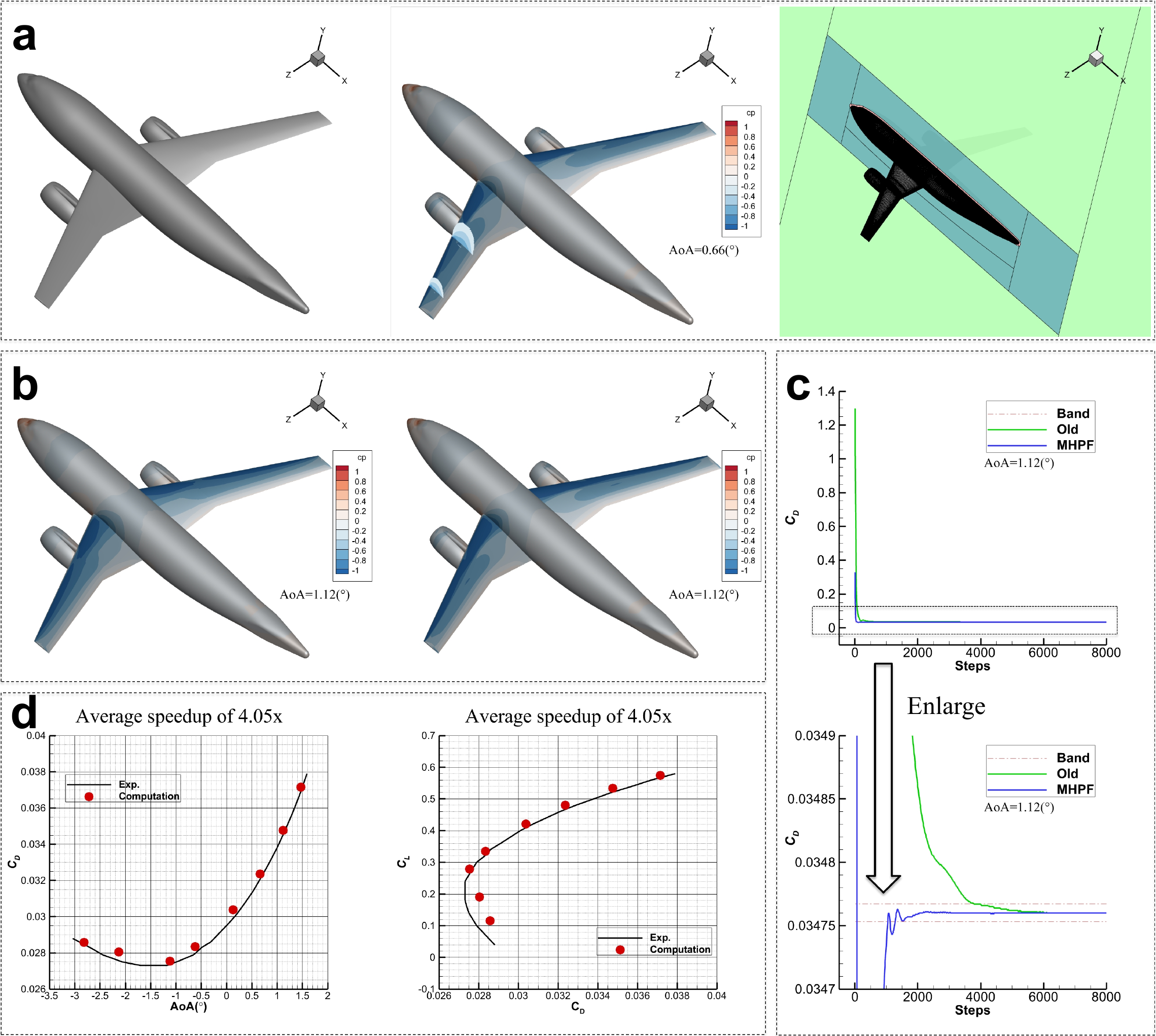}
    \caption{\textbf{DLR-F6-WBNP computational geometry and comparison of computational results.}
a, Computational geometry, representative flow field and flow-field regions on grids at different layers. The computational grid contains 148 blocks and 14,380,032 cells, with $y^+\approx1$. b, Comparison between the predicted ( left ) and converge ( right ) surface-pressure contours at $\alpha=1.12^\circ$. The predicted flow again shows high similarity to the converged flow, with only minor differences downstream of the pylon. c, Comparison and local enlargement of drag convergence curves obtained using conventional CFD and MHLF at the representative angle of attack. d, Comparison of drag coefficient and lift-to-drag ratio obtained using MHLF with wind-tunnel measurements. The results agree closely except at two small angles of attack. This discrepancy is consistent with the fact that the experimental model involved forced transition and a certain degree of elastic deformation during wind-tunnel testing, which causes differences between measured aerodynamic forces and most fully turbulent numerical simulations \citep{laflin2004summary}.}
    \label{fig:DLR-F6-WBNP}
\end{figure}

The DLR-F6-WBNP configuration, consisting of wing, body, nacelle and pylon components, was released by the AIAA Applied Aerodynamics Technical Committee (APATC) for the second Drag Prediction Workshop. The case is used mainly to assess the ability of CFD methods to simulate aerodynamic interference among realistic aircraft components \citep{wang2018overview}. The configuration uses a modern wide-body aircraft wing with sweep and spanwise twist. Compared with the smooth DLR-F6 configuration, it includes an engine nacelle and pylon. Flow separation may occur near the nacelle-pylon junction, and interactions among the nacelle, pylon and wing can alter the shock position and boundary-layer state. This configuration therefore places strong demands on CFD solvers for simulating boundary-layer separation and reattachment, as well as shock-boundary-layer interaction. It has substantial practical value and has become a commonly used transonic CFD benchmark.

We used the same procedure described above to perform numerical simulations for the training conditions. The training conditions were $Ma=0.75$ and $Re=3.10\times10^{6}$, with $\alpha=-3^\circ \sim 1.5^\circ$ at intervals of $0.25^\circ$. The governing equations were the full Navier--Stokes equations, and the turbulence model was the two-equation SST model. The MHLF framework was then trained on the resulting flow fields using the same training configuration as in the X-38 case. The comparison conditions were again excluded from the training set. The tested angles of attack were $-2.82^\circ, -2.13^\circ, -1.12^\circ, -0.62^\circ, 0.13^\circ, 0.66^\circ, 1.12^\circ, 1.47^\circ$. The computational geometry and comparison results are shown in Fig.~\ref{fig:DLR-F6-WBNP}.

\begin{table}[!htbp]
\centering
\caption{
Convergence performance and computational cost for the DLR-F6-WBNP case.
}
\label{tab:dlrf6_wbnp_steps_time}
\renewcommand{\arraystretch}{1.15}
\setlength{\tabcolsep}{1pt}
\begin{tabular}{lcccccccccc}
\toprule
\multirow{2}{*}{Method} &
\multicolumn{8}{c}{Convergence steps $N_c$ at AoA $\alpha$ (deg)} &
\multirow{2}{*}{Total steps} &
\multirow{2}{*}{Total time (s)} \\
\cmidrule(lr){2-9}
& $-2.82^\circ$ & $-2.13^\circ$ & $-1.12^\circ$ & $-0.62^\circ$
& $0.13^\circ$ & $0.66^\circ$ & $1.12^\circ$ & $1.47^\circ$
& & \\
\midrule
Old       & 5040 & 5280 & 5184 & 4800 & 4320 & 2532 & 4812 & 6012 & 37980 & 63996.3 \\
PointCNN           & $\times$ & $\times$ & $\times$ & $\times$ & $\times$ & $\times$ & $\times$ & $\times$ & $\times$ & $\times$ \\
PointConv          & 9504 & 9528 & 4800 & 6924 & 7164 & 8184 & 7056 & 6636 & 59796 & 100756.3 \\
PointMLP           & $\times$ & $\times$ & $\times$ & $\times$ & $\times$ & $\times$ & $\times$ & $\times$ & $\times$ & $\times$ \\
PointNet++          & 5136 & 5496 & 6312 & 5808 & 5100 & 4476 & 4008 & 4764 & 41100 & 69253.5 \\
PointNeXt          & 5976 & 7308 & 6672 & 7332 & 6840 & 4464 & 3240 & 5112 & 46944 & 79100.6 \\
PointTransformerV3 & 5760 & 5904 & 5964 & 5820 & 4884 & 3732 & 3984 & 5892 & 41940 & 70668.9 \\
\textbf{MHLF (ours)}               & $\mathbf{2340}^{1}$ & $\mathbf{2484}^{1}$ & $\mathbf{2868}^{1}$ &
                     $\mathbf{1824}^{2}$ & $\mathbf{2448}^{3}$ & $\mathbf{1440}^{3}$ &
                     $\mathbf{1524}^{4}$ & $\mathbf{1512}^{4}$ &
                     $\mathbf{16440}$ & $\mathbf{15787.1}$ \\
\bottomrule
\end{tabular}

\vspace{2mm}
\begin{flushleft}
\footnotesize
Superscripts 1, 2, 3 and 4 denote the 10-3-2, 10-3-3, 10-3-4 and 10-3-5 modes, respectively.
\end{flushleft}
\end{table}

Table~\ref{tab:dlrf6_wbnp_steps_time} compares MHLF with conventional CFD and baseline models. As in the previous case, the baseline models generally show poor predictive capability for flow fields, and all require longer computational times than conventional CFD. By contrast, MHLF reduces the total number of computational steps across the eight comparison conditions to 43.3\% of that required by conventional CFD. After the speed-up from asynchronous iteration is included, MHLF achieves an average acceleration of 4.05 in this case, with the largest speed-up of 6.44 at $\alpha=1.47^\circ$. These results show that, for a transonic case with a more complex geometry, MHLF can still predict flow phenomena involving interactions between shock waves and boundary layers in multi-component configurations with good accuracy.

\subsection{TRAP Wing case}

\begin{figure}[!htbp]
    \centering
    \includegraphics[width=\textwidth]{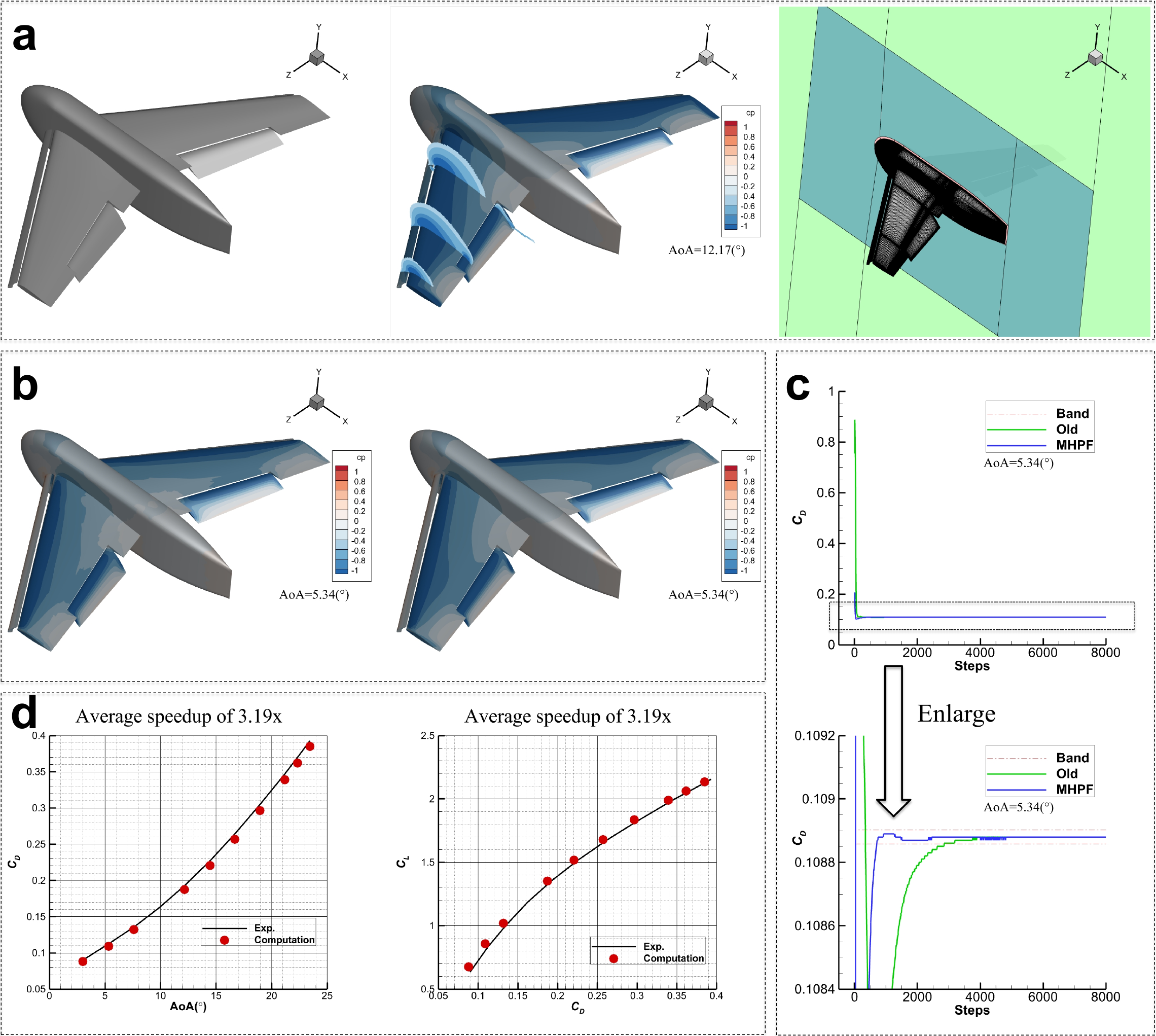}
    \caption{\textbf{TRAP Wing computational geometry and comparison of computational results.}
a, Computational geometry, representative flow field and flow-field regions on grids at different layers. The computational grid contains 253 blocks and 16,494,848 cells, with $y^+\approx1$. b, Comparison between the predicted ( left ) and converge ( right ) surface-pressure contours at $\alpha=5.34^\circ$. Apart from slight non-smooth artifacts, the predicted flow is highly similar to the converged flow. c, Comparison and local enlargement of drag convergence curves obtained using conventional CFD and MHLF at the representative angle of attack. d, Comparison of drag coefficient and lift-to-drag ratio obtained using MHLF with wind-tunnel measurements. The two sets of results agree closely across the full comparison range.}
    \label{fig:TRAP-WING}
\end{figure}

To comprehensively evaluate the predictive capability of MHLF, we further tested a more complex subsonic case after completing the supersonic and transonic comparisons. Subsonic aircraft typically have more control surfaces and are geometrically much more complex than high-speed aircraft. We therefore selected a representative high-lift configuration for validation. Flow fields around high-lift configurations are usually closely associated with complex physical phenomena such as flow separation, transition and boundary-layer mixing, all of which are difficult for CFD.

The high-lift configuration tested here is the part-span configuration of the Trap Wing benchmark released at the first High Lift Prediction Workshop (HiLiftPW-1) in 2010. This configuration consists of a large-chord, moderate-aspect-ratio, three-element system mounted on a simplified fuselage, including a leading-edge slat, main wing and trailing-edge flap \citep{yuntao2018overview}. Compared with the full-span case released at the same workshop, this configuration is closer to a realistic civil aircraft. The trailing-edge flap has a scissors-like deflection, which makes grid generation difficult using conventional abutting-grid methods. In this work, the case was generated using an patched grid method.

As described above, the training conditions were $Ma=0.15$ and $Re=1.5\times10^{7}$, with $\alpha=3^\circ \sim 24^\circ$ at intervals of $0.5^\circ$. The governing equations were the full Navier--Stokes equations, and the turbulence model was the one-equation SA model. Except that the number of training epochs was increased to $1000$, the model training configuration was the same as in the X-38 case. The comparison angles of attack were $3.01^\circ$, $5.34^\circ$, $7.61^\circ$, $12.17^\circ$, $14.44^\circ$, $16.70^\circ$, $18.95^\circ$, $21.20^\circ$, $22.33^\circ$ and $23.46^\circ$. The computational geometry and comparison results are shown in Fig.~\ref{fig:TRAP-WING}.

\begin{table}[!htbp]
\centering
\caption{
Convergence performance and computational cost for the TRAP Wing case.
}
\label{tab:trap_wing_steps_time}
\renewcommand{\arraystretch}{1.15}
\setlength{\tabcolsep}{1pt}

\begin{tabular}{lcccccccccccc}
\toprule
\multirow{2}{*}{Method} &
\multicolumn{10}{c}{Convergence steps $N_c$ at AoA $\alpha$ (deg)} &
\multirow{2}{*}{Total steps} &
\multirow{2}{*}{Total time (s)} \\
\cmidrule(lr){2-11}
& $3.01^\circ$ & $5.34^\circ$ & $7.61^\circ$ & $12.17^\circ$ & $14.44^\circ$
& $16.70^\circ$ & $18.95^\circ$ & $21.20^\circ$ & $22.33^\circ$ & $23.46^\circ$
& & \\
\midrule
Old       & 3780 & 3408 & 3768 & 3312 & 2796 & 2712 & 2652 & 2652 & 2808 & 2424 & 30312 & 18338.8 \\
PointCNN           & 7716 & 6540 & 7212 & 6780 & 6612 & 6996 & 6084 & 3816 & 4848 & 6828 & 63432 & 38376.4 \\
PointConv          & 5148 & 4800 & 3540 & 5424 & 4740 & 7068 & 7908 & 7596 & 8136 & 10296 & 64656 & 39116.9 \\
PointMLP           & 3984 & 4512 & 2556 & 4140 & 2376 & 3612 & 2616 & 2592 & 3456 & 2448 & 32292 & 19536.7 \\
PointNet++          & 3312 & 2688 & 4476 & 2724 & 2436 & 2952 & 1836 & 4764 & 4524 & 2928 & 32640 & 19747.2 \\
PointNeXt          & 6276 & 5508 & 7128 & 5664 & 5340 & 5112 & 5376 & 3540 & 3924 & 8460 & 56328 & 34078.4 \\
PointTransformerV3 & 4788 & 4980 & 6924 & 1500 & 2316 & 1452 & 4068 & 4488 & 3576 & 8112 & 42204 & 25533.4 \\
\textbf{MHLF (ours)}               & $\mathbf{3180}^{1}$ & $\mathbf{840}^{2}$ & $\mathbf{2640}^{2}$ &
                     $\mathbf{1488}^{2}$ & $\mathbf{1368}^{3}$ & $\mathbf{1740}^{3}$ &
                     $\mathbf{1056}^{3}$ & $\mathbf{1380}^{3}$ & $\mathbf{1020}^{4}$ &
                     $\mathbf{960}^{5}$ & $\mathbf{15672}$ & $\mathbf{5752.0}$ \\
\bottomrule
\end{tabular}

\vspace{2mm}
\begin{flushleft}
\footnotesize
Superscripts 1, 2, 3, 4 and 5 denote the 10-3-1, 10-3-3, 10-3-5, 10-3-7 and 10-3-9 modes, respectively.
\end{flushleft}
\end{table}

Table~\ref{tab:trap_wing_steps_time} compares MHLF with conventional CFD and baseline models. As in the preceding cases, the baseline models generally show poor predictive capability for flow fields, and all require longer computational times than conventional CFD. For two of the baseline models, the predicted initial conditions more than double the convergence time. We attribute this behavior to the high geometric complexity of the configuration. The control surfaces create multiple flow-field regions, producing flow separation at different spatial locations and substantial differences in flow morphology across regions. Under these conditions, a non-hierarchical model has difficulty overcoming feature masking among regions and accurately predicting the local flow state in each region. MHLF retains a clear advantage over the baseline models even for this complex geometry. Across the ten evaluation conditions, the total number of computational steps required by MHLF is 49.6\% of that required by conventional CFD. After the speed-up from asynchronous iteration is included, the average acceleration reaches 3.19, with the largest speed-up of 7.01 at $\alpha=5.34^\circ$. This case also shows that, although MHLF uses hierarchical processing, its predictive capability remains limited when highly complex geometries induce strongly heterogeneous multi-region flow states. A highly adaptive dynamic regional training strategy should therefore be considered in future work.

\section{Discussion}

Compared with conventional CFD acceleration strategies, MHLF represents a new CFD paradigm for engineering-scale flow simulation. Conventional methods mainly seek to accelerate convergence by improving the numerical algorithm itself. MHLF instead combines faster numerical correction with a deep learning improvement of the initial flow field, thereby shortening the convergence trajectory. A simple distance-speed analogy clarifies this difference. If the distance from point A to point B is $s$ and the moving speed is $v$, then the travel time is $t=s/v$. To reduce $t$, conventional methods mainly increase $v$, corresponding to the convergence speed of the CFD solver. MHLF acts on both terms. It increases the effective update efficiency through hierarchical asynchronous iteration, and reduces $s$ through initial-field prediction, which decreases the gap between the initial and converged flow fields. The present results suggest that the efficiency gain from hierarchical asynchronous iteration alone is constrained by grid-distribution characteristics and may be limited in the short term. By contrast, initial-field prediction provides a larger acceleration margin. In some comparison conditions, MHLF reduced the number of convergence steps to approximately $1/6 \sim 1/8$ of that required by conventional initialization. Further improvements in model architecture and regional learning may extend this acceleration to a broader range of conditions.

This advantage is obtained at very low additional cost. The time and computational resources required to generate learning data on the coarsened training grids are almost negligible, making MHLF highly valuable for the large numbers of flow states encountered in engineering CFD applications. Moreover, other flow-field prediction methods often exhibit large random errors that are difficult to correct, whereas the final flow-field results obtained with MHLF are numerically accurate after CFD correction. This property makes the framework particularly important for high-precision data applications, especially in aerospace engineering. In this work, the coarsening stride is 4, so the flow-field data are compressed to $1/64$ of the original size. However, the multiresolution representation based on multigrid can compress training data much further. We have also implemented four consecutive coarsening operations with strides of 3, 3, 2 and 2, compressing aircraft flow-field data with 10.4 billion grid cells to $1/46656$ of the original data size while preserving all major macroscopic flow structures. Large-scale learning of engineering-scale flow-field data based on MHLF should therefore no longer be fundamentally limited by the original grid size or hardware capacity.

Although MHLF shows substantial promise in the three canonical cases, several issues remain to be addressed. First, flow states vary markedly with angle of attack, and separation regions and shock structures continue to evolve. The random sampling strategy used for sparse point clouds also makes model predictions insufficiently stable, with high accuracy in some states and weaker performance in others. Improving the model to achieve consistently high-quality flow-field prediction across all states will be a focus of future work. Second, more extensive testing is still needed for more practical geometries and flow structures, as well as for the generalization capability of the prediction model beyond the training set. Third, current network training is still dominated by data fitting and does not sufficiently incorporate physical constraints such as flux conservation and boundary conditions during learning. The relationship between deep learning prediction and physical consistency may therefore be critical for further improving such methods. Finally, the present cases are validated using structured grids, although MHLF is also compatible with unstructured-grid methods. The implementation of hierarchical iteration on unstructured grids has been described in prior work on hierarchical iteration methods for CFD numerical solution \citep{meng2026hierarchical}. However, owing to the inherent advantages of structured grids, grid coarsening and flow-field reconstruction are relatively straightforward to implement in the present framework. For unstructured grids, the same regional hierarchy and geometric-multigrid-based multiresolution prolongation and reconstruction framework could in principle be used, but the implementation is more involved. The performance of MHLF for unstructured-grid methods therefore remains to be further confirmed.

Looking ahead, a more urgent application is to establish the ability of prediction models to generalize across aircraft geometries. In the present cases, the computational geometries are fixed. In engineering CFD, however, many simulations require accurate prediction of flow fields around deformed configurations. Achieving this capability would provide a new solution pathway for full-envelope aircraft performance evaluation, aircraft design optimization, and aeroelastic problems involving static and dynamic coupling between aerodynamics and structures. The long and highly repetitive workload associated with conventional CFD would be greatly reduced, and MHLF-based aircraft design, optimization and evaluation systems could achieve a step change in efficiency. We will address these issues and challenges progressively in future work.

\section{Methods}
\begin{figure}[!htbp]
    \centering
    \includegraphics[width=\textwidth]{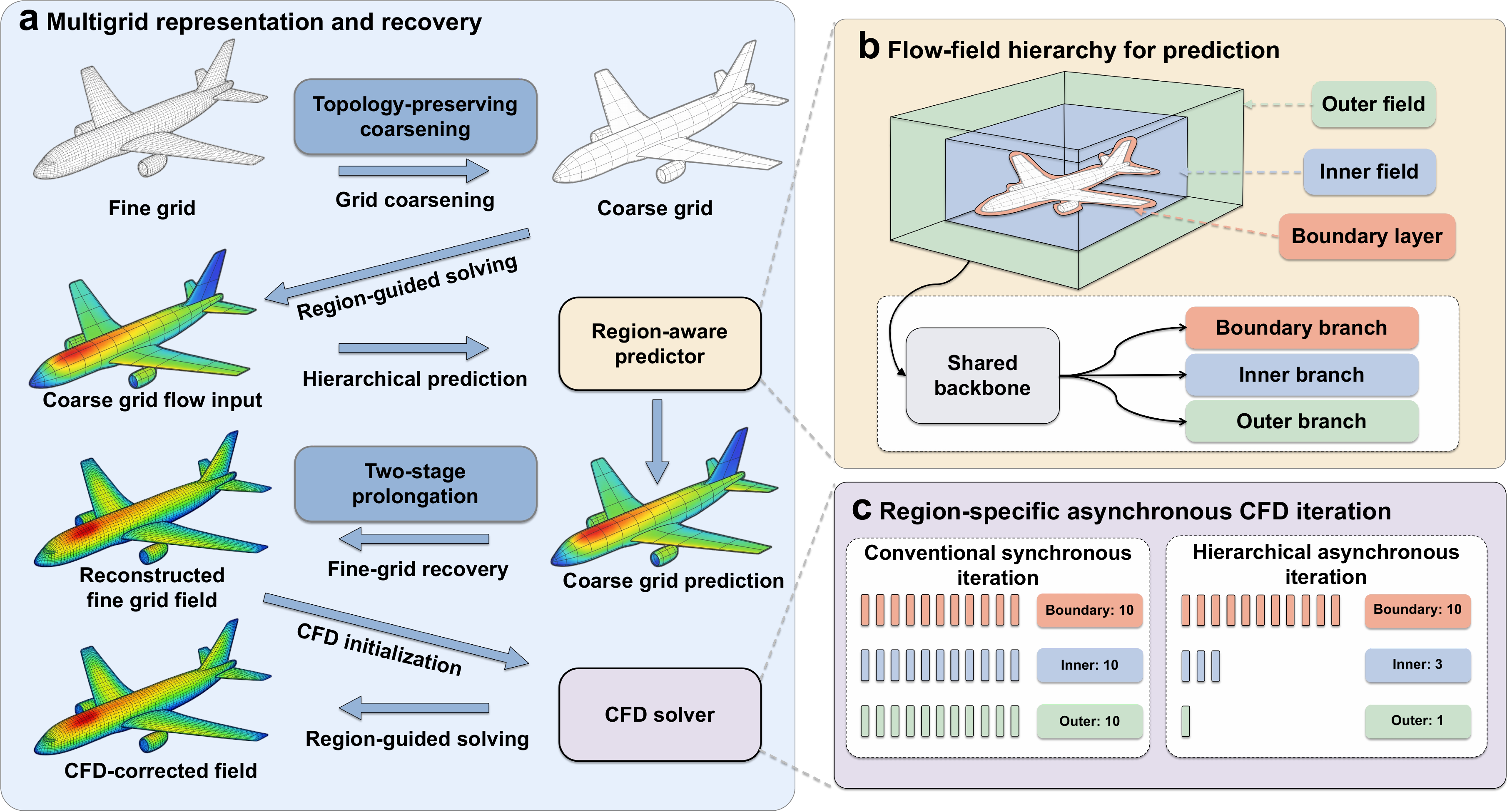}
    \caption{\textbf{Overview of the MHLF framework.}
a, Multigrid representation and reconstruction. The original fine grid is converted into a topologically consistent coarse-grid representation through a topology-preserving coarsening strategy, and flow-field prediction is performed on this coarse grid. The predicted coarse-grid flow field is then reconstructed on the fine grid through two-stage prolongation and used as the initial condition for the CFD solver.
b, Flow-field hierarchy for prediction. The flow field is organized into boundary-layer, inner-field and outer-field regions, and is modeled using a shared backbone network with region-specific branches for region-aware prediction.
c, Region-specific asynchronous CFD iteration. The reconstructed fine-grid flow field is further corrected by CFD, with an asynchronous iteration strategy that assigns different iteration counts to different flow-field regions.}
    \label{fig:framework}
\end{figure}

Figure~\ref{fig:framework} shows the overall workflow of MHLF. MHLF decomposes direct engineering-scale full-field prediction into four coupled steps that consist of coarse-grid representation learning, region-aware flow-field prediction, fine-grid reconstruction and CFD correction. First, the original fine grid is mapped to a low-resolution representation space through topology-preserving coarsening, reducing the computational complexity of large-scale flow-field learning. The computational domain is then organized into boundary-layer, inner-field and outer-field regions according to flow-field characteristics. The model performs adaptive prediction for these regions using a shared backbone network and region-specific branches. The coarse-grid prediction is reconstructed on the original fine grid through two-stage prolongation and used as the initial flow field for CFD correction. Finally, the same hierarchy guides asynchronous iteration scheduling during CFD correction, allowing the boundary-layer, inner-field and outer-field regions to use different iteration counts according to their convergence characteristics and thereby improving computational efficiency.

In MHLF, the multigrid method provides cross-resolution representation and reconstruction, whereas the hierarchical strategy supports both region-adaptive learning and region-specific asynchronous numerical iteration. By integrating coarse-grid learning, fine-grid reconstruction and subsequent CFD correction into a unified workflow, MHLF provides an effective approach for predicting engineering-scale three-dimensional flow fields while retaining high-fidelity numerical accuracy.

\subsection{Multigrid representation and reconstruction}
\begin{figure}[!htbp]
    \centering
    \includegraphics[width=\textwidth]{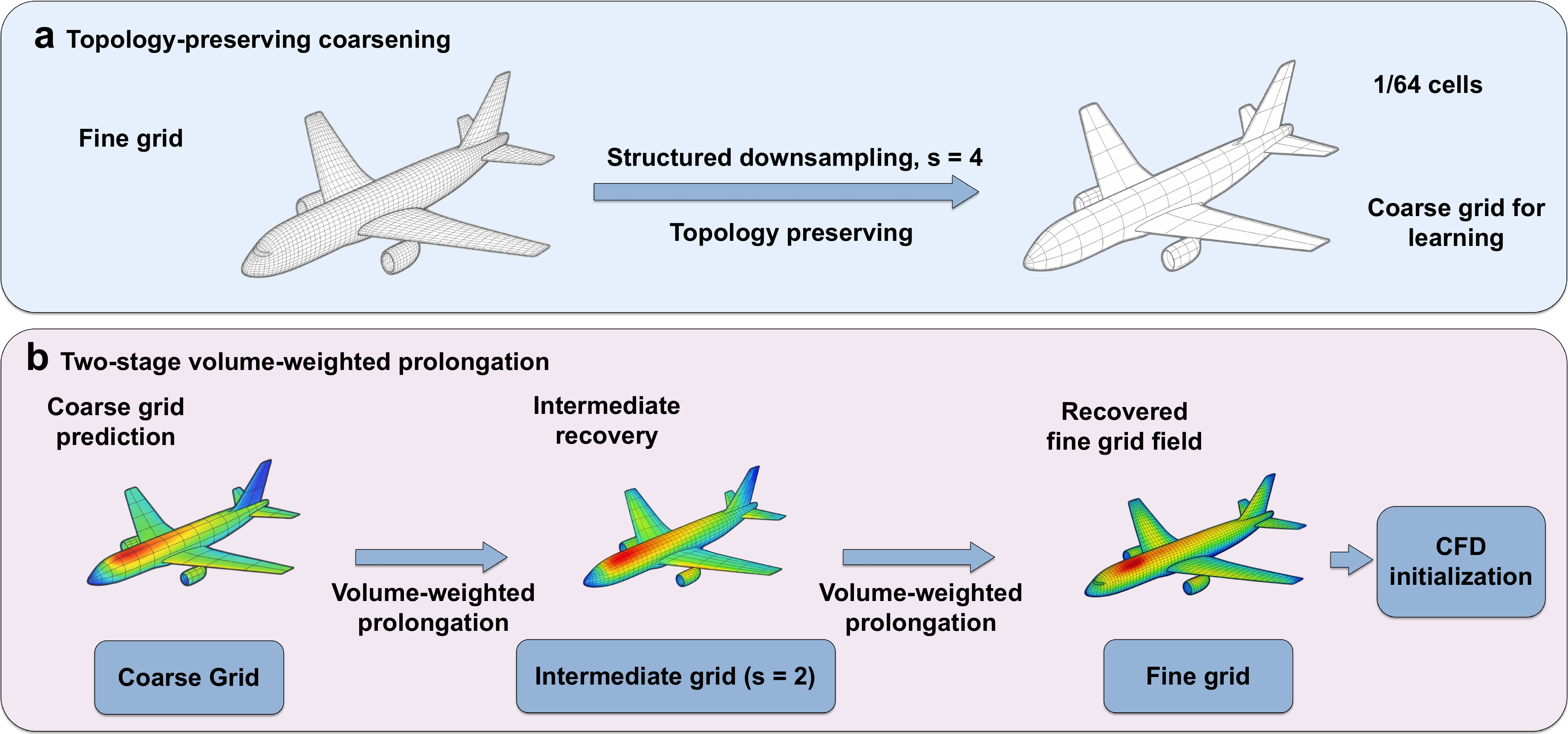}
    \caption{\textbf{Multigrid representation and reconstruction in MHLF.}
a, Topology-preserving coarsening. The original fine grid is structurally coarsened with stride $s=4$ to construct a topologically consistent coarse-grid representation for learning, reducing the number of grid cells to approximately $1/64$ of that in the fine grid.
b, Two-stage volume-weighted prolongation. The predicted coarse-grid flow field is first mapped to an intermediate grid through volume-weighted prolongation, and is then reconstructed on the original fine grid through a second volume-weighted prolongation step, yielding a high-resolution flow field for CFD initialization.}
    \label{fig:multigrid}
\end{figure}

As shown in Fig.~\ref{fig:multigrid}, the multigrid module of MHLF consists of a topology-preserving grid-coarsening process that constructs the coarse-grid representation and a two-stage volume-weighted prolongation process that reconstructs the predicted coarse-grid flow field on the original high-resolution grid. The central idea is to perform the main learning task in a low-resolution but geometrically consistent representation space, and then map the low-resolution prediction back to the fine grid to provide a high-resolution initial flow field for subsequent CFD correction.

\subsubsection{Topology-preserving grid coarsening}
We construct the cross-resolution representation using geometric multigrid (GMG). Here, GMG is used to define the learning space rather than only to accelerate numerical iteration. The aim is to reduce the size of the regression problem while preserving the grid correspondence needed for later reconstruction on the CFD grid.

Let the original high-resolution structured grid be denoted as $G_{\mathrm{fine}}$. A coarse grid is obtained by sampling the grid with a fixed stride along each computational direction:
\[
G_{\mathrm{coarse}}
=
\left\{
x_{i,j,k}\in G_{\mathrm{fine}}
\;\middle|\;
i,j,k \equiv 0 \; (\mathrm{mod}\; s)
\right\},
\]
where $s$ is the coarsening stride. In this work, $s=4$. One point is retained for every four points along each grid edge, and the computational grid is generated from the coarsened grid edges. In three dimensions, this reduces the number of coarse-grid cells to $1/64$ of the original fine-grid count.

Because $G_{\mathrm{coarse}}$ is obtained by direct coarsening from $G_{\mathrm{fine}}$, the two grids remain consistent in geometry and spatial topology. This correspondence links low-cost prediction to subsequent fine-grid CFD correction. We perform CFD simulations on $G_{\mathrm{coarse}}$ to generate supervised learning data. Although the coarse-grid solution cannot resolve all local details, it preserves the large-scale flow structures and main physical distributions needed for initialization.

\subsubsection{Two-stage volume-weighted prolongation}

CFD correction requires the predicted field to be reconstructed on the original high-resolution grid $G_{\mathrm{fine}}$. The coarse-grid prediction must therefore be reconstructed in the fine-grid space before CFD correction. Because the resolution differs by a factor of $4$ along each spatial direction and the total number of cells differs by a factor of $64$, direct interpolation can distort regions with strong grid stretching or large local volume variation.
To reduce the error caused by a large single-step mapping, we introduce a medium-resolution grid $G_{\mathrm{medium}}$ between $G_{\mathrm{coarse}}$ and $G_{\mathrm{fine}}$. Reconstruction is decomposed into two prolongation steps, $G_{\mathrm{coarse}} \rightarrow G_{\mathrm{medium}}$ and $G_{\mathrm{medium}} \rightarrow G_{\mathrm{fine}}$. The medium grid is obtained by structurally coarsening the fine grid with stride $s=2$, so it remains topologically consistent with both the coarse and fine grids. This intermediate level reduces the resolution jump in each mapping and stabilizes reconstruction.

At each reconstruction level, a volume-weighted prolongation operator $I_{\mathrm{vol}}(\cdot)$ maps the low-resolution flow field to a higher-resolution grid. The operator accounts for local control-volume differences instead of relying only on geometric distance. It constructs interpolation weights from the volume distribution of cells in the target neighborhood, which helps reduce distortion near boundaries and locally refined regions. The reconstruction process is
\[
\widehat{\mathbf{Y}}_{\mathrm{medium}}
=
I_{\mathrm{vol}}
\left(
\widehat{\mathbf{Y}}_{\mathrm{coarse}},
G_{\mathrm{medium}}
\right),
\]
\[
\widehat{\mathbf{Y}}_{\mathrm{fine}}
=
I_{\mathrm{vol}}
\left(
\widehat{\mathbf{Y}}_{\mathrm{medium}},
G_{\mathrm{fine}}
\right).
\]

The final reconstructed flow field $\widehat{\mathbf{Y}}_{\mathrm{fine}}$ is used as the initial condition for CFD correction. This step converts coarse-grid learning into a CFD acceleration pathway.

\subsection{Flow-field hierarchy for prediction and CFD correction}
The flow-field hierarchy makes regional physical differences explicit before prediction and CFD correction. As shown in Figs.~\ref{fig:partition-net} and~\ref{fig:framework}c, this part of MHLF includes hierarchy definition with region-wise normalization, region-aware prediction on flow-field regions, and region-specific asynchronous CFD iteration. This organization avoids a single homogeneous treatment of boundary-layer, inner-field and outer-field regions.

\subsubsection{Flow-field hierarchy and region-wise normalization}

\begin{figure}[!htbp]
    \centering
    \includegraphics[width=\textwidth]{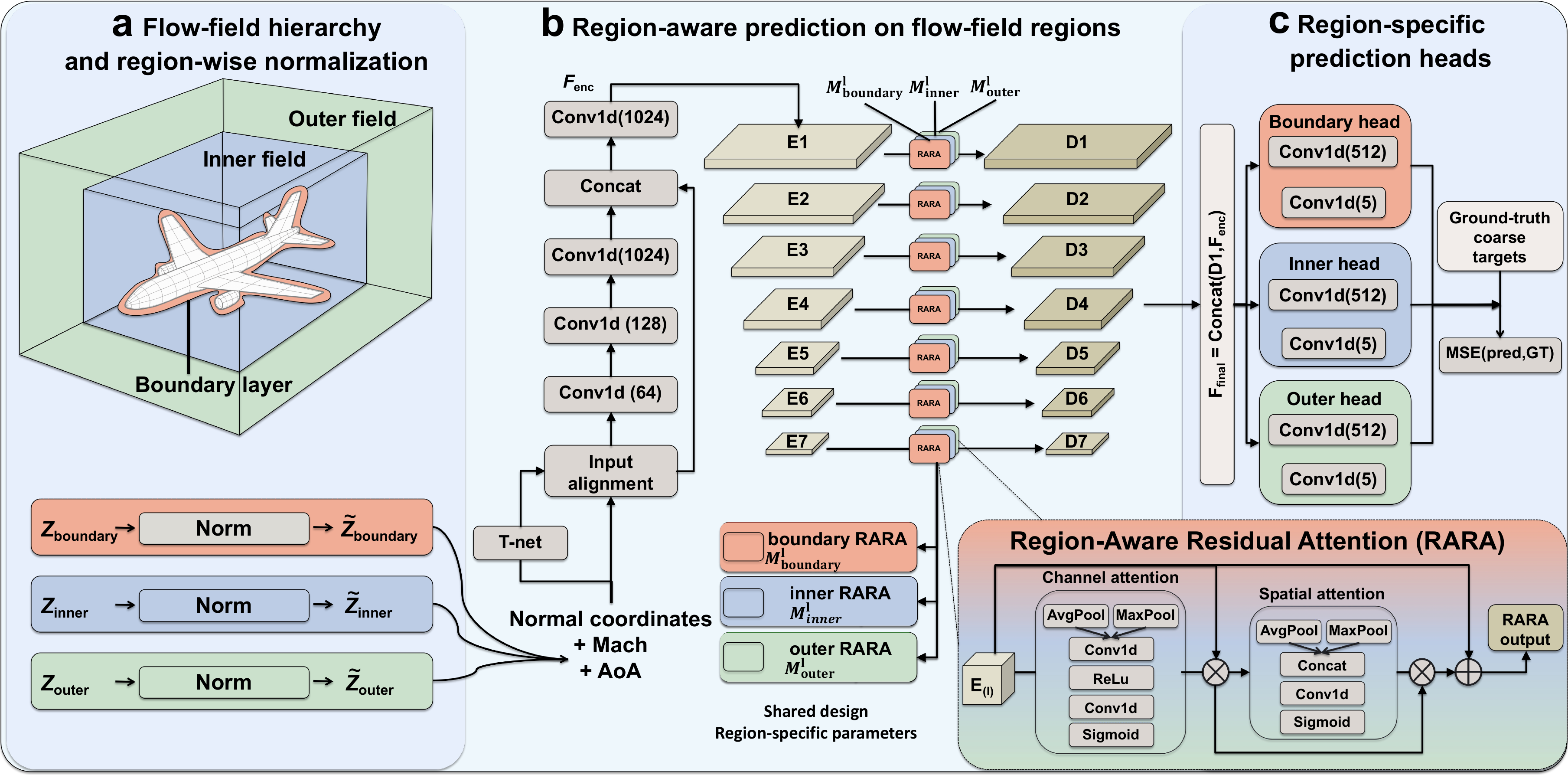}
    \caption{\textbf{Flow-field hierarchy and region-aware prediction in MHLF.}
a, Flow-field hierarchy and region-wise normalization. The flow field is organized into boundary-layer, inner-field and outer-field regions, and each region is normalized independently to reduce scale imbalance and feature masking among heterogeneous flow regions.
b, Region-aware prediction on flow-field regions. Normalized coordinates and flow-condition parameters are processed by a shared backbone network, while region-specific RARA branches adaptively modulate multiscale features in different physical regions.
c, Region-specific prediction heads. The final latent representations are mapped to coarse-grid flow variables by independent regression heads corresponding to the boundary-layer, inner-field and outer-field regions.}
    \label{fig:partition-net}
\end{figure}

As shown in Fig.~\ref{fig:partition-net}a, the flow field is organized into a boundary-layer region $\Omega_{\mathrm{boundary}}$, an inner-field region $\Omega_{\mathrm{inner}}$ and an outer-field region $\Omega_{\mathrm{outer}}$. This hierarchy separates regions with different statistical scales and numerical stiffness. The region $\Omega_{\mathrm{boundary}}$ lies near the wall and contains strong grid stretching and steep local gradients. The region $\Omega_{\mathrm{inner}}$ lies between the near-wall region and the far field and contains transitional, medium-scale flow structures. The region $\Omega_{\mathrm{outer}}$ is dominated by smoother flow variations.

For any region
$
r \in \{\Omega_{\mathrm{boundary}},\Omega_{\mathrm{inner}},\Omega_{\mathrm{outer}}\}$,
let $N_r$ denote the number of sampled points in this region. The geometric coordinates and flow variables are denoted by
\[
\mathbf{X}_r =(x,y,z)\in \mathbb{R}^{N_r\times 3},\qquad
\mathbf{Y}_r =(\rho,u,v,w,p)\in \mathbb{R}^{N_r\times 5}.
\]

Because spatial scales and variable distributions differ across regions, each region is normalized independently rather than by a global transform. For the $i$th point in region $r$,
\[
\tilde{Z}^{(i)}_{r}
=
\frac{Z^{(i)}_{r}-z^{\min}_{r}}
{z^{\max}_{r}-z^{\min}_{r}},
\qquad
z^{\min}_{r}=\min_{i} Z^{(i)}_{r},\;
z^{\max}_{r}=\max_{i} Z^{(i)}_{r},
\]
where $\mathbf{Z}_r$ can represent either the regional coordinates $\mathbf{X}_r$ or the flow variables $\mathbf{Y}_r$. This region-wise normalization prevents near-wall geometry from being compressed by global scaling and balances the input distributions used for region-aware prediction.

\subsubsection{Region-aware prediction on flow-field regions}

As shown in Fig.~\ref{fig:partition-net}b,c, prediction is performed on the flow-field regions within the coarse-grid representation. The network uses a Backbone--Branch Modulation architecture to share global aircraft-flow information while allowing region-specific responses. The shared backbone extracts geometric and flow representations common to all regions. The branches modulate skip connections and output mapping for each flow-field region.

For any region $r$, the network input consists of normalized coordinates and global flow-condition parameters. After the Mach number $Ma$ and angle of attack $\alpha$ are broadcast to all points in the region, the input tensor is
\[
\mathbf{U}_r
=
\left[
\tilde{\mathbf{X}}_r,\;
Ma,\;
\alpha
\right]
\in \mathbb{R}^{N_r\times 5},
\]
where $\tilde{\mathbf{X}}_r\in\mathbb{R}^{N_r\times 3}$ denotes the normalized regional coordinates. A lightweight T-Net first aligns the input to reduce sensitivity to geometric and flow-condition variations:
\[
\mathbf{U}^{\mathrm{ali}}_r=\mathbf{U}_r\mathbf{T}_r,
\]
where $\mathbf{T}_r\in\mathbb{R}^{5\times 5}$ is the input alignment matrix. The aligned input $\mathbf{U}^{\mathrm{ali}}_r$ is mapped by pointwise convolution to shallow local features $\mathbf{F}_{\mathrm{local}}\in\mathbb{R}^{N_r\times 64}$, and successive pointwise convolutions extract deep features $\mathbf{F}_{\mathrm{deep}}\in\mathbb{R}^{N_r\times 1024}$. The two feature tensors are concatenated and compressed to preserve local geometric information before multiscale encoding:
\[
\mathbf{F}_{\mathrm{enc}}
=
\phi_{\mathrm{red}}
\big(
\mathrm{Concat}(\mathbf{F}_{\mathrm{deep}},\mathbf{F}_{\mathrm{local}})
\big)
\in\mathbb{R}^{N_r\times 1024}.
\]

The feature $\mathbf{F}_{\mathrm{enc}}$ then enters a seven-level encoder path, producing multiscale features $\{\mathbf{E}^{(l)}\}_{l=1}^{7}$, which are progressively reconstructed in a mirrored decoder path. In a standard U-Net, feature fusion at level $l$ is written as
\[
\mathbf{D}^{(l)}
=
\mathcal{D}^{(l)}
\left(
\mathrm{Concat}(\mathbf{D}^{(l+1)},\mathbf{E}^{(l)})
\right).
\]
To prevent skip connections from applying the same feature emphasis to regions with different flow characteristics, we introduce the RARA module at each level. For each level $l$, three modules, $\mathcal{M}^{(l)}_{\mathrm{boundary}}$, $\mathcal{M}^{(l)}_{\mathrm{inner}}$ and $\mathcal{M}^{(l)}_{\mathrm{outer}}$, are constructed, and the branch is selected according to the input region:
\[
\mathbf{E}^{(l)}_{r,\mathrm{rara}}
=
\mathcal{M}^{(l)}_{r}\big(\mathbf{E}^{(l)}\big).
\]

Given the input feature $\mathbf{E}^{(l)}\in\mathbb{R}^{N_r\times C_l}$, RARA first applies channel attention:
\[
\mathbf{a}^{(l)}_{r,c}
=
\sigma\!\left(
\mathrm{MLP}(\mathrm{AvgPool}_N(\mathbf{E}^{(l)}))
+
\mathrm{MLP}(\mathrm{MaxPool}_N(\mathbf{E}^{(l)}))
\right)
\in\mathbb{R}^{1\times C_l},
\]
where $\sigma(\cdot)$ is the Sigmoid function. This gives the channel-recalibrated feature
\[
\mathbf{E}^{(l)}_{r,c}
=
\mathbf{E}^{(l)}\odot \mathbf{a}^{(l)}_{r,c}.
\]
Spatial attention is then applied:
\[
\mathbf{a}^{(l)}_{r,s}
=
\sigma\!\left(
\mathrm{Conv}_{7}
\left(
\mathrm{Concat}
\big(
\mathrm{AvgPool}_C(\mathbf{E}^{(l)}_{r,c}),
\mathrm{MaxPool}_C(\mathbf{E}^{(l)}_{r,c})
\big)
\right)
\right)
\in\mathbb{R}^{N_r\times 1},
\]
The resulting feature is
\[
\mathbf{E}^{(l)}_{r,s}
=
\mathbf{E}^{(l)}_{r,c}\odot \mathbf{a}^{(l)}_{r,s}.
\]

Finally, RARA outputs the region-modulated feature in residual form:
\[
\mathbf{E}^{(l)}_{r,\mathrm{rara}}
=
\mathbf{E}^{(l)}+\mathbf{E}^{(l)}_{r,s}.
\]

Thus, the feature fusion of the level-$l$ decoder becomes
\[
\mathbf{D}^{(l)}
=
\mathcal{D}^{(l)}
\left(
\mathrm{Concat}
\big(
\mathbf{D}^{(l+1)},\mathbf{E}^{(l)}_{r,\mathrm{rara}}
\big)
\right).
\]

At the end of the decoder, the top-level decoded feature is concatenated again with the encoder-entry feature to obtain the final representation:
\[
\mathbf{F}_{\mathrm{final}}=\mathrm{Concat}(\mathbf{D}^{(1)},\mathbf{F}_{\mathrm{enc}}).
\]
Independent prediction heads $\mathcal{H}_{\mathrm{boundary}}$, $\mathcal{H}_{\mathrm{inner}}$ and $\mathcal{H}_{\mathrm{outer}}$ keep the final regression consistent with the regional organization. For any region $r$, the prediction is
\[
\widehat{\mathbf{Y}}_r
=
\mathcal{H}_r(\mathbf{F}_{\mathrm{final}}).
\]
Each prediction head consists of two pointwise convolution layers, first compressing the feature to 512 dimensions and then mapping it to five physical variables.

During training, the mean squared error is computed separately for each region and then combined as a weighted sum:
\[
\mathcal{L}
=
\sum_{r\in\{\Omega_{\mathrm{boundary}},\Omega_{\mathrm{inner}},\Omega_{\mathrm{outer}}\}}
\lambda_r\,
\mathrm{MSE}\big(\widehat{\mathbf{Y}}_r,\tilde{\mathbf{Y}}_r\big),
\]
where $\lambda_r$ is the regional loss weight. Unless otherwise specified, $\lambda_r=1$ in this work.

The prediction model outputs the five non-conservative variables, density, pressure and the three velocity components. The turbulence variables required by the CFD solver are supplemented from coarse-grid turbulence fields at a nearby angle of attack. This keeps the learning target focused on the main flow variables while retaining the turbulence information needed for CFD correction.

\subsubsection{Region-specific asynchronous CFD iteration}

This step uses the fine-grid flow field obtained by two-stage prolongation as the initial condition and obtains a physically constrained converged solution through subsequent iteration. We further introduce a hierarchical iteration strategy in this process \citep{meng2026hierarchical}, as briefly described below.

As shown in Fig.~\ref{fig:framework}c, we no longer apply a unified synchronous update to the entire flow field. Instead, asynchronous iteration is performed according to the physical characteristics and convergence behaviour of each region. Figure~\ref{fig:framework}c illustrates a commonly used 10-3-1 iteration strategy, which can be dynamically adjusted for different computational conditions. For mainstream time-marching CFD solvers based on local time stepping, the boundary-layer region usually requires more iterations because of its small grid scale, steep gradients and stronger numerical stiffness. The inner-field region has larger grid scales and smaller flow gradients, so the number of iterations can be moderately reduced. The outer-field region exhibits smoother flow variations and can use fewer iteration steps. This asynchronous iteration strategy assigns different iteration counts to different regions. It does not change the governing equations or discretization framework of the solver, but substantially reduces redundant updates in smooth regions by adjusting the iteration strategy region by region. This enables balanced advancement of the full flow field and markedly improves numerical iteration efficiency.

\subsection{CFD simulation method}

The CFD solver provides the reference converged fields for supervised learning and the correction step used after MHLF initialization. All simulations were performed using NNW-FSI, the fluid-structure interaction software of the National Numerical Wind Tunnel platform developed by the China Aerodynamics Research and Development Center. The CFD module is based on structured grids and discretizes the Reynolds-averaged Navier--Stokes (RANS) equations in generalized coordinates using the finite-volume method, as shown in the following equation. Convective terms are discretized by the MUSCL-Roe upwind scheme, and diffusive terms by a second-order central scheme. The turbulence models include the one-equation SA model and the two-equation SST model. The discrete equations are solved mainly with LU-SGS. Preconditioning is used for low-speed calculations, and geometric multigrid and massively parallel computing are used to accelerate convergence. Matching grids are used for geometrically simpler cases, whereas patched grids are used for complex configurations to handle scissor gaps caused by control-surface deflection.
\[
\frac{\partial \hat{Q}}{\partial \tau}
+
\frac{\partial \hat{E}}{\partial \xi}
+
\frac{\partial \hat{F}}{\partial \eta}
+
\frac{\partial \hat{G}}{\partial \zeta}
=
\frac{\partial \hat{E}_v}{\partial \xi}
+
\frac{\partial \hat{F}_v}{\partial \eta}
+
\frac{\partial \hat{G}_v}{\partial \zeta}
.
\]
Here, $\hat{Q}$ denotes the vector of conservative variables. $\hat{E}$, $\hat{F}$ and $\hat{G}$ denote the inviscid fluxes in curvilinear coordinates, and $\hat{E}_v$, $\hat{F}_v$ and $\hat{G}_v$ denote the corresponding viscous fluxes. These are mature mainstream CFD algorithms and are therefore not described further. The multigrid method used in this work uses three grid levels to support coarsening with $s=4$.

\subsubsection{Convergence criterion}

To accurately compare the computational efficiency of CFD, it is essential to precisely count number of iterations to convergence. Since aerodynamic forces in complex flows often do not converge directly to a fixed value but instead exhibit small-amplitude oscillations, and given that there is currently no widely accepted unified convergence criterion for such flows in the CFD community, we adopted convergence criterion proposed for hierarchical CFD iteration \citep{meng2026hierarchical}. Details are provided in Appendix 1.

\subsubsection{Initial-condition dependence}

Because MHLF changes the CFD initial condition, we assessed whether the final converged solution depends on the starting field. Such dependence can occur in separated or discontinuous flows. It is well documented in high-lift configurations, where cold starts and hot starts near stall can converge to different states. The AIAA High Lift Prediction Workshop therefore recommended using the converged flow field from a nearby lower angle of attack as the initial condition near stall.

The present numerical experiments show the following. First, for all attached flows and flows with small separation bubbles, both uniform and predicted initial fields converge to the same solution. Second, when large surface separation is present, the converged results obtained from uniform and predicted initial fields are also identical in most states. In a small number of states, slight differences may occur, usually with aerodynamic-force differences below 1\textperthousand. Near the critical angles of attack where separation first appears or is about to disappear, however, differences in surface-separation morphology can cause this discrepancy to exceed 1\%. This situation corresponds to the coexistence of multiple attractors in a nonlinear dynamical system, meaning that the system has multiple stable solutions.

\begin{figure}[!htbp]
    \centering
    \includegraphics[width=\textwidth]{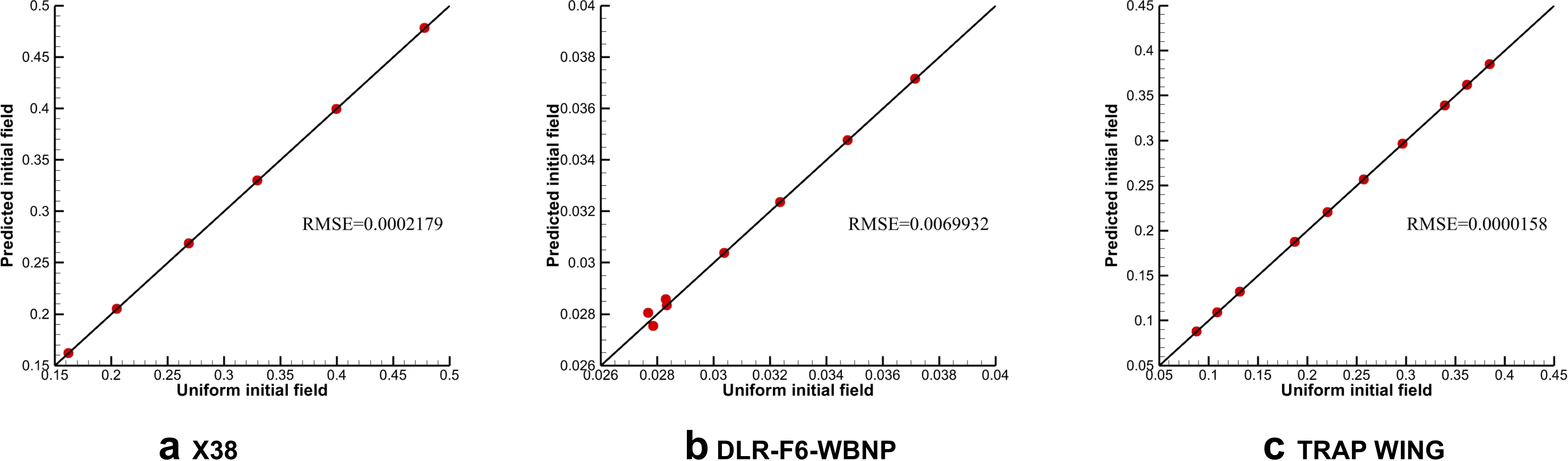}
    \caption{\textbf{Comparison of drag coefficients obtained from different initial conditions.}
a--c, Comparison of drag coefficients obtained from uniform and predicted initial conditions for the X-38, DLR-F6-WBNP and TRAP Wing cases, respectively. The horizontal axis denotes the drag coefficient obtained from the uniform initial condition, and the vertical axis denotes the drag coefficient obtained from the predicted initial condition. RMSE denotes the root mean squared relative error between the two sets of results.}
    \label{fig:start}
\end{figure}

In practical CFD applications, when different initial conditions yield inconsistent results and grid or solver changes do not remove the discrepancy, the solution initialized closer to the final flow state is commonly selected. Following this practice, when uniform and predicted initial conditions produced different results, we reported the solution obtained from the predicted initial condition. Figure~\ref{fig:start} compares the two sets of results. Differences mainly occur near critical separation states, such as the negative-angle-of-attack states of the WBNP configuration, whereas other states are almost identical. Initial condition dependence is not the focus of this study and is not discussed further.

\subsection{Baseline models}

To test whether the performance gains come from the multigrid-hierarchical formulation rather than generic point-based model capacity, we selected six representative point-based learning models: PointCNN \citep{li2018pointcnn}, PointConv \citep{wu2019pointconv}, PointMLP \citep{ma2022rethinking}, PointNet++ \citep{qi2017pointnet++}, PointNeXt \citep{qian2022pointnext} and Point Transformer V3 \citep{wu2024point}. These baselines cover major paradigms for irregular point-cloud learning, including point convolution, hierarchical set abstraction, MLP-based learning and Transformer-based attention. This set enables a systematic comparison between MHLF and classical or recent point-cloud learning methods.

We further compared the prediction error and model size of all methods across the three benchmark cases. As shown in Table~\ref{tab:mse_params_comparison}, MHLF achieves the lowest MSE on all three datasets while using the fewest trainable parameters. This result indicates that the improved prediction performance of MHLF does not come from a larger model capacity, but from the multigrid representation and flow-field hierarchy.
\begin{table}[!htbp]
\centering
\caption{Prediction error and model size across benchmark cases.}
\label{tab:mse_params_comparison}
\renewcommand{\arraystretch}{1.15}
\setlength{\tabcolsep}{8pt}
\begin{tabular}{lcccc}
\toprule
\multirow{2}{*}{Model} & \multicolumn{3}{c}{MSE} & \multirow{2}{*}{Parameters} \\
\cmidrule(lr){2-4}
& X-38 & DLR-F6-WBNP & TRAP Wing & \\
\midrule
PointCNN           & $2.0133\times10^{-3}$ & $1.8514\times10^{-3}$ & $4.5090\times10^{-5}$ & 11,348,113 \\
PointConv          & $2.5870\times10^{-3}$ & $1.4144\times10^{-3}$ & $5.2769\times10^{-5}$ & 13,399,335 \\
PointMLP           & $6.1433\times10^{-4}$ & $1.0361\times10^{-3}$ & $1.8280\times10^{-4}$ & 15,307,549 \\
PointNet++         & $4.2203\times10^{-3}$ & $2.4633\times10^{-3}$ & $4.1228\times10^{-5}$ & 15,965,317 \\
PointNeXt          & $7.0552\times10^{-3}$ & $3.3955\times10^{-3}$ & $1.6759\times10^{-4}$ & 11,118,365 \\
PointTransformerV3 & $1.8946\times10^{-2}$ & $5.9227\times10^{-3}$ & $2.0732\times10^{-4}$ & 46,150,597 \\
\textbf{MHLF (ours)} & $\mathbf{2.2128\times10^{-4}}$ & $\mathbf{2.4497\times10^{-4}}$ & $\mathbf{2.8135\times10^{-5}}$ & $\mathbf{9,374,100}$ \\
\bottomrule
\end{tabular}
\end{table}

\section{Data availability}
Owing to institutional confidentiality regulations, the data that support the findings of this study are not publicly available. All data are available from the corresponding author upon reasonable request.

\section{Code availability}
Owing to institutional confidentiality regulations, the code used in this study is not publicly available. All code is available from the corresponding author upon reasonable request.

\bibliography{sn-bibliography}

\section{Convergence criterion}

\begin{figure}[!htbp]
    \centering
    \includegraphics[width=\textwidth]{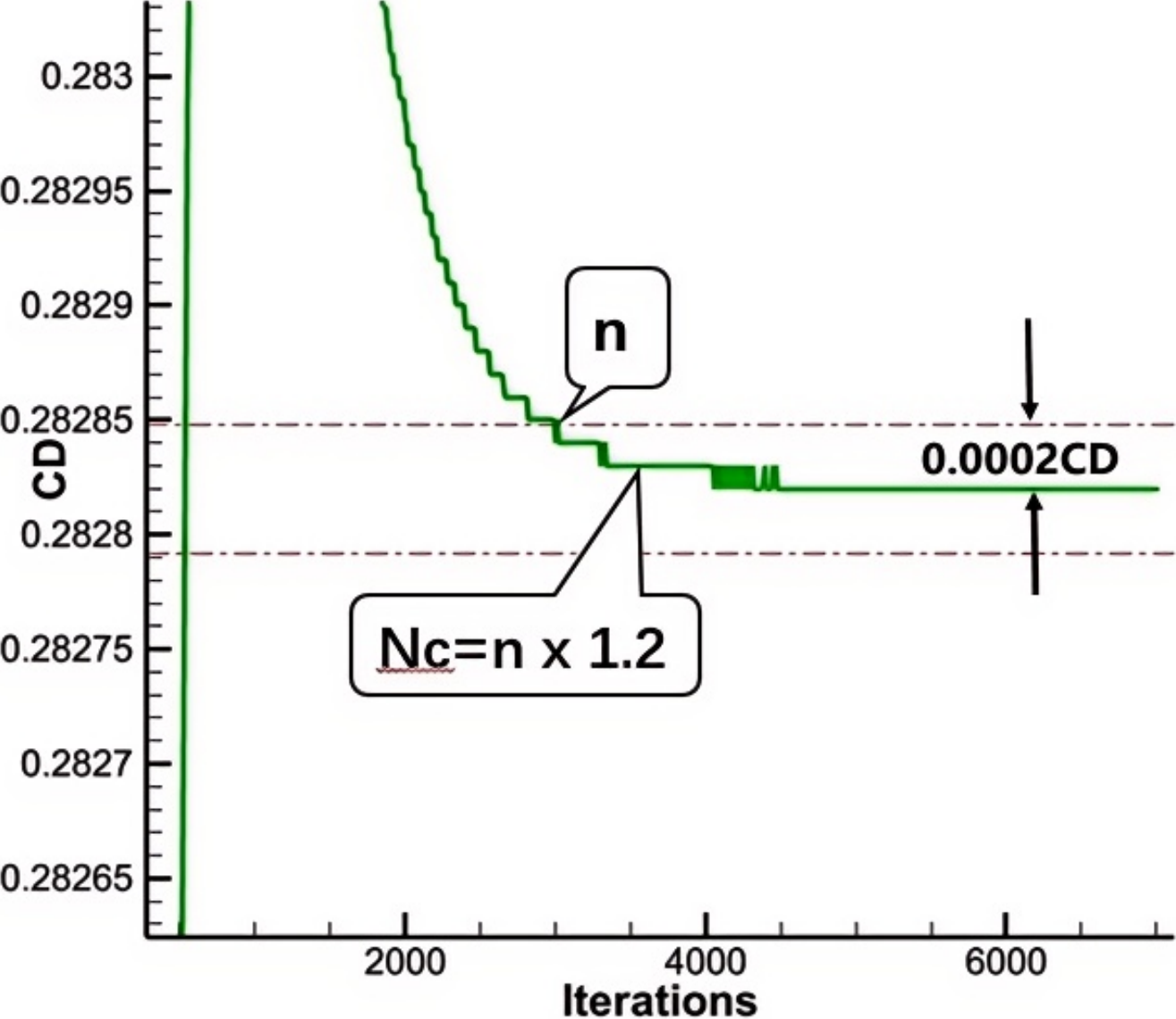}
    \caption{\textbf{Drag-coefficient-based convergence criterion.}
The final converged drag coefficient $C_D$ is used as the reference value, and the interval $C_D \pm 0.0002C_D$ is defined as the convergence band. Let $n$ denote the iteration step at which the drag-coefficient convergence history enters this band for the last time. The convergence step count is then defined as $N_c = 1.2n$. For cases in which the drag coefficient exhibits small-amplitude fluctuations and does not converge to a strictly fixed value, the reference value is taken as the mean drag coefficient during the statistically converged stage, and $n$ is determined after filtering and smoothing.}
    \label{fig:convergence criterion}
\end{figure}

To quantitatively compare the convergence efficiency of different CFD strategies, this work adopts the convergence criterion proposed in the study on hierarchical iteration methods for CFD numerical solution, as illustrated in Fig.~\ref{fig:convergence criterion}. In the reference study, the convergence band was set to $\pm 0.0001 C_D$. Here, we relax this threshold to $\pm 0.0002 C_D$, mainly because the CFD software used in this work outputs aerodynamic coefficients with five significant digits. For computational states with more complex geometries and spatial flow structures, truncation error can cause fluctuations in the last digit of the aerodynamic output. In this situation, a convergence band of $\pm 0.0001 C_D$ is comparable to the uncertainty of the numerical output, making the drag-coefficient history prone to oscillations near the convergence band and making the convergence step count difficult to quantify. We therefore use a slightly wider convergence band to improve the robustness and practicality of the criterion. This work does not address whether this criterion is the optimal definition of CFD convergence. Its purpose is to provide an objective, reproducible and operational standard for quantifying convergence steps, so that the convergence efficiency of different initial conditions and iteration strategies can be compared within the same numerical framework.


\end{document}